\documentstyle[sprocl,epsf]{article}

\newcommand{\eqa}{\begin{eqnarray}}
\newcommand{\ena}{\end{eqnarray}}
\newcommand{\eqar}{\begin{array}}
\newcommand{\enar}{\end{array}}
\newcommand{\eqn}{\begin{equation}}
\newcommand{\enq}{\end{equation}}

\font\blackboard=msbm10 scaled \magstep1
\font\blackboards=msbm7
\font\blackboardss=msbm5
\newfam\black
\textfont\black=\blackboard
\scriptfont\black=\blackboards
\scriptscriptfont\black=\blackboardss
\def\Bbb#1{{\fam\black\relax#1}}
\def\Bbb{\bf}
\def\yboxit#1#2{\vbox{\hrule height #1 \hbox{\vrule width #1
\vbox{#2}\vrule width #1 }\hrule height #1 }}
\def\fillbox#1{\hbox to #1{\vbox to #1{\vfil}\hfil}}
\def\ybox{{\lower 1.3pt \yboxit{0.4pt}{\fillbox{8pt}}\hskip-0.2pt}}
\def\comments#1{}

\def \AdS {{\rm AdS}}
\def \IIa {IIa}
\def \IIb {IIb}

\def \Tr {{\rm Tr}}

\def\BR{\Bbb{R}}

\def\p{\partial}

\def\ket#1{|#1\rangle}
\def\vev#1{\langle{#1}\rangle}

\def\CN{{\cal N}}

\begin{document}

\rightline{IC/99/7}

\rightline{RU-99-06}

\title{TWO LECTURES ON AdS/CFT CORRESPONDENCE}

\author{MICHAEL R. DOUGLAS\footnote{
Supported in part by DOE grant DE-FG05-90ER40559.}}

\address{Department of Physics and Astronomy, Rutgers University,\\
Piscataway, NJ 08855, USA\\
and\\
I.H.E.S., Le Bois-Marie, Bures-sur-Yvette France 91440}

\author{S. RANDJBAR-DAEMI}
\address{The Abdus Salam International Centre for Theoretical Physics,\\
Trieste, Italy}

%addresses:
% 1 Dept. of Physics and Astronomy\\
 %Rutgers University\\
% Piscataway, NJ 08855
% 2 I.H.E.S., Le Bois-Marie, Bures-sur-Yvette France 91440
 %3 I.C.T.P.

\maketitle\abstracts{
This is a write-up of two lectures on AdS/CFT correspondance
given by the authors at the 1998 Spring School at the Abdus Salam ICTP.
}

\section{Introduction}

The best candidate theories of quantum gravity, maximal supergravity and
superstring theory, are formulated in more than four dimensions.
Traditionally four-dimensional physics is derived using the Kaluza-Klein
approach, which assumes space-time is a product of a non-compact manifold
$M$ and a compact manifold $K$.  Einstein's equations in the full space-time
then relate the curvature of these manifolds: if $M$ is to be flat
Minkowski space, $K$ must be Ricci-flat (or related to a Ricci-flat
manifold in a simple way).
Massless fields on $M$ arise as zero modes of differential operators on $K$.

These models are easily studied for $K$ an $n$-torus, and candidate
$K$'s exist with reduced supersymmetry.  However their
Ricci-flat metrics are not known explicitly and studying them
requires tricky, indirect methods.
Rather, the
next simplest possibility is to take $K$ to be a homogeneous space
such as the $n$-sphere $S^n$.  The isometries of $K$ lead to vector
fields on $M$, so these compactifications produce gauged supergravity.

These models were much studied in the 1980's \cite{duff};
solutions exist but the
positive curvature of $K$ forces $M$ to have equal and opposite negative
curvature, making them at first sight phenomenologically irrelevant.
The simplest possibility is $M$ of constant negative curvature, i.e.
anti-de Sitter space $\AdS_d$.

These models also have maximal supersymmetry and thus are ``too good''
not to find a place in the web of theories known as M theory.
A clear application emerged with the study of extremal black hole
solutions \cite{bhsusy}.  Such solutions of supergravity in Minkowski
space can preserve up to half of the global supersymmetry,
but it was observed that in the near horizon limit they often preserve
maximal supersymmetry.
This comes about because the near horizon geometry is equivalent
to an $\AdS$ compactification.

Black hole physics underwent a revolution with the advent of the
$D$-brane as one could now find configurations with the same long range
fields as an extremal black hole, but whose microscopic degrees of
freedom are known explicitly.  This led to the first generally accepted
computation of
the Bekenstein-Hawking entropy of a black hole, by Strominger amd
Vafa \cite{strvafa}.
Certain extremal black holes in type \IIb\  superstring theory
compactified on $T^5$ can be represented as bound states of D$1$ and
D$5$-branes wrapped on the torus -- both objects have exactly the
same long-range fields, and are identified in string theory.
On the other
hand the D-brane system can also be described in world-volume terms --
from open string theory one derives a two-dimensional conformal field
theory, for which the entropy can be computed exactly.
This entropy, certain Hawking emission rates,
and many other observables agree exactly between the two descriptions.

The essential reason that conformal field theory appears in the discussion
is to reflect the additional near-horizon symmetry visible in the space-time
description.  By D-brane considerations to be reviewed below,
the near horizon limit in space-time is reflected
in the low energy limit of the world-volume theory, which takes it to
a fixed point with conformal symmetry.  Such a $d$-dimensional conformal
theory will then have $SO(d,2)$ space-time symmetry.
This symmetry enhancement is exactly parallel to
the enhancement of the black hole symmetry to the
isometry group of $\AdS_{d+1}$, $SO(d,2)$.
The same relation holds in the superconformal case.

Another example of this type
in type \IIb\  superstring theory is the D$3$-brane.
A configuration of $N$ coincident D$3$-branes in Minkowski space
is identified with another extremal black hole,
as we discuss below.  On the other hand,
its low energy world-volume theory
is $N=4$ four-dimensional supersymmetric Yang-Mills theory.
Thus we might conjecture that the entropy of this extremal black hole
is equal to the entropy of $N=4$ Yang-Mills theory.

Thus it was something of a surprise when it was found that -- in distinction
to the D$1$--D$5$ black hole -- these entropies, computed in
conventional perturbation theory, did not agree \cite{gkp}.
Further work found many quantities that do not agree, even in the D$1$--D$5$
system -- for example, the metric as seen by a probe brane \cite{dps,mprobe}.

Of course the simplest resolution of all such issues
is to argue that conventional perturbation theory is not accurate for these
questions.   In fact the gauge theory
limit which must reproduce supergravity is large $N$
and large 't Hooft coupling $g^2 N$, where perturbation theory fails,
and not much was known.
So how does one proceed ?

Recently Maldacena has reversed this identification in a fruitful way
\cite{maldacena}.
Instead of trying to derive the properties of a black hole from the
D-brane theory, one makes a precise conjecture stating that the D-brane
theory is {\it equivalent to}
the black hole and indeed all of the gravitational
dynamics needed to describe it.
Assuming this conjecture,
one can derive results for the large 't Hooft coupling
limit of gauge theory, by
doing computations in $\AdS$ supergravity.
Even better, by simple modifications of the background geometry,
one can get results in gauge theories with reduced or no supersymmetry.

This relation has the same spirit as previously conjectured
dualities: here
there exists a single theory which reduces to perturbative gauge theory for
weak coupling and to supergravity for strong coupling.
There are a few observables constrained by supersymmetry to be
independent of the coupling, and the basic test
of the hypothesis is that these agree.  Explicit computation has
revealed that additional quantities agree in the two limits;
no compelling statement has yet been made
as to what should be expected to agree and why.

On the other hand the entropy and indeed almost all observables
are expected to disagree\footnote{
The agreement for the D$1$-D$5$ system turns out to depend on a result
special to two-dimensional conformal field theory; the invariance of
central charge under marginal deformations.
}
and thus one interprets these as
non-trivial functions of the coupling for which one
now has results in both limits.
In general, one has as yet no direct way to test these
predictions from either gauge theory or string theory;
however even qualitative agreement can be
regarded as evidence.  In the case of the D$3$-brane, large $N$
four-dimensional gauge theory was much studied over the years and
numerous guesses made for its behavior.
As it turns out the predictions
from AdS/CFT contradict some of these guesses, but in ways which appear
to form a new consistent picture of large $N$ $\CN=4$ gauge theory;
the better motivated guesses (e.g. for a deconfinement transition in
pure Yang-Mills at finite temperature) are confirmed.

The aim of these lectures was to introduce this subject
at an elementary level. We shall mostly be concerned with the case of the
D$3$ brane in type \IIb\  theory.

\section{The $3$-brane of type \IIb}

The massless sector of \IIb\ string theory is \IIb\ supergravity,
which is described in many references including \cite{diverse,gsw}.
On general grounds a $p$-brane will be a source of a $p+1$-form
gauge potential and thus the minimal subsector of the theory required
to describe a $3$-brane will be the metric $g_{MN}$
and the four-form potential $C$.  This potential has a self-dual
five-form field strength $F$ for which the action is somewhat complicated but
for our purposes we will only need the equations of motion
\eqa
{1\over \kappa^2} R_{MN} = {1\over 6} F_{MIJKL} F_N^{IJKL} \cr
F = * F; \qquad dF = 0
\ena
and supersymmetry transformations of the gravitino
\eqn\label{susy}
\delta \psi_M =
(D_M + {i \kappa\over 5!} F_{ABCDE} \Gamma^{ABCDE} \Gamma_M) \xi
\enq
where $\xi$ and $\psi_M$ are complex ten-dimensional Weyl spinors
(representing the sum of two Majorana-Weyl spinors), and $\kappa$ is
the ten-dimensional Newton's constant.

We write the $3$-brane ansatz
\eqn
ds^2= e^{2A(y)} dx^\mu dx^\nu \eta_{\mu\nu} + e^{-2B(y)} dy^2
\enq
and
\eqn
F = (1+*) d e^{4C(y)} dx^0 dx^1 dx^2 dx^3
\enq
where
$\eta_{\mu\nu}$ , $0\le \mu, \nu \le 3$ is the usual $3+1$
Minkowski metric, while $y^i$ parameterize the six dimensions
transverse to the brane.  Ultimately the functions $A$, $B$ and $C$ will
be taken to depend only on $r=|y|$ to describe a single $3$-brane.

The brane solution will preserve supersymmetries corresponding to
parameters $\xi$ for which the gravitino variation is zero.
We can decompose $\xi$ into the product of four and six dimensional
spinors of each chirality
\eqn
 \xi = f_+(y) \xi^{(4+)} \epsilon^{(6+)} +
	f_-(y) \xi^{(4-)} \epsilon^{(6-)}
\enq
and substituting the ansatz into (\ref{susy}) only $\Gamma^{0123}$ appears,
so up to $1/2$ the bulk supersymmetry can be preserved, say the
components $\xi^{(4+)} \epsilon^{(6+)}$ giving the equivalent of
$N=4$ supersymmetry in $d=4$.  Requiring (\ref{susy}) to have solutions we
can derive the relation $A=B=C$ and substituting this into $dF=0$
we find that $e^{-4C} \equiv f(y)$ must satisfy Laplace's equation in
six flat transverse dimensions.

The final three-brane solutions are
\eqn\label{threebrane}
ds^2= f^{-1/2} dx^\mu dx^\nu \eta_{\mu\nu} + f^{1/2} dy^2
\enq
and
\eqn\label{Ffield}
F = (1+*) d(f^{-1}) dx^0 dx^1 dx^2 dx^3
\enq
with
\eqn
f(y) = 1 + \sum_{i=1}^n {Q_i\over |y-y_i|^4} .
\enq

The charges $Q_i$ are quantized as one can see by an extension of
the Dirac argument: consider a second three-brane which extends in
$x^0$ and (say) $y^4$, $y^5$ and $y^6$; from its point of view (\ref{Ffield})
is a magnetic gauge potential, and by moving it in the three
remaining transverse dimensions we can rerun Dirac's argument.
There is actually a subtle factor of two coming from the self-duality
of $F$ but the final quantization condition is
\eqn\label{chargeq}
Q_i = 4 \pi g_s N_i l_s^4
\enq
with $N_i$ integer.

Let us take $n=1$ and $y_1=0$ to get a solution symmetric under
$SO(6)$ rotations of $y$.  Define $r=|y|$.
The parameter $g_s N$ appearing in (\ref{chargeq}) will play a central role,
so we define $\lambda = 4 \pi g_s N$ and write
\eqn\label{ffunction}
f(y) = 1 + {\lambda l_s^4\over r^4} .
\enq
On general grounds in \IIb\ string theory
we believe the supergravity description of this solution only
for $r >> l_s$, because at shorter distances we cannot justify
ignoring the massive closed string modes.
This is good as brane solutions in supergravity are typically
singular (the extreme three-brane is an exception) but leads us to ask:
How do we describe the regime $r << l_s$ ?

\section{The $3$-brane as a D$3$-brane}

As is well-known by now,
we can define the {\it same} solution of \IIb\ string theory
by introducing boundaries in our world-sheet theory, constrained
to live on the plane $y=0$.  The charge $N$ is represented by
allowing an $N$-valued Chan-Paton factor; the multi-center solutions
correspond to allowing several types of boundaries ending on the planes
$y=y_i$.
We can compute the long-range supergravity fields around this plane at
leading order in $g_s$ by doing world-sheet path integrals on a disk
with a graviton or RR vertex operator inserted;
we can also verify that the Dirichlet boundary conditions linearly relate
the left and right-moving world-sheet operators generating space-time
supersymmetry as
\eqn
Q_L = \Gamma^{0123} Q_R
\enq
and thus preserve the same supersymmetries as the supergravity solution
(\ref{threebrane}).
These considerations lead us to {\it identify} the D-brane as the unique
object in string theory
corresponding to the field configuration (\ref{threebrane}).

This description is non-singular and in contrast to the supergravity
description becomes simpler as we consider short transverse distance
scales $r << l_s$.  Fluctuations of the D-brane are described by
exciting open strings ending on the D-brane, and effects described by
open strings stretched to the radius $r$ will be associated with the
mass scale $m = T_s r = r/2\pi l_s^2$ (the usual string tension energy).
In general excited open strings will also contribute but if $r << l_s$
these will be much heavier than the lightest open strings.

Thus we can simplify our brane theory by taking a scaling limit of
short distances and large string tension:
$r\rightarrow 0$ and $l_s \rightarrow 0$, and work in the energy units
set by the stretched strings: $u = r / l_s^2$ fixed.
This is the low energy limit of the world-volume theory,
which is $N=4$ $U(N)$ super Yang-Mills theory (SYM)
with gauge coupling $g^2_{YM} = 4\pi g_s$.\footnote{
Our conventions are such that S-duality is $g_s\rightarrow 1/g_s$.
}

We review some well-known facts about this theory.  The $N=4$ gauge
multiplet contains besides the gauge field six real adjoint scalars $Y^i$
and four Majorana gauginos $\chi^{I\alpha}$.  The action can be
obtained by dimensional reduction from $N=1$, $d=10$ SYM and contains
the potential $V = \sum_{i<j} [Y^i,Y^j]^2$.  It respects an $SO(6)$
R symmetry which is simply inherited from the original rotational
symmetry in $d=10$ SYM and acts on $Y^i$ as a vector.

As a field theory on $\BR^{3,1}$, the moduli space
of supersymmetric vacua is parameterized by commuting matrices
$[Y^i,Y^j]=0$ up to gauge transformation; i.e. is $\BR^{6N}/S_N$.
This is identical to the parameter space for the most general
harmonic function (\ref{ffunction})
with all $Q_i=1$ so the moduli space is identical
in the two descriptions.

If we take all $Y^i=0$ we have an unbroken conformal invariance
$SO(4,2)$ in the classical theory and (unusually) even after quantization,
because the beta function vanishes.  This directly implies scale invariance
and indirectly implies conformal invariance in the quantum theory.\footnote{
Actually, when we are taking the low energy limit, any starting
theory will flow to a conformal theory.  Examples in which the theory
only becomes conformal in this limit include the D$1$-D$5$ black
hole, the $2$-brane in M theory, and many others.}
In fact we have a superconformal
quantum theory in four dimensions.  Now the superconformal algebra is
generated by the product of the conformal and supersymmetry algebras,
but is larger -- we can apply conformal inversion to the
supercharges $Q^{I\alpha}$, to obtain partners $S^{I\alpha}$, leading
to a total of $32$ fermionic generators.  The full algebra is $SU(2,2|4)$
and given in \cite{superspace}. We shall come back to this algebra in section
5 and give a brief descrption of some of its
 unitary irreducible representations.

Clearly this is a very interesting
point in the moduli space, however it was never too
clear what its physics might
be.  In particular, the usual particle and S-matrix
interpretation for quantum field theory is problematic as Green's functions
are not expected to have the required analytic structure.

On the other hand there are numerous qualitative consequences of conformal
invariance which are well known in two dimensions and which we can expect
to hold here.  One of these is the operator-state correspondence.
This is most clearly formulated by using radial quantization: we choose
a point, say $x^\mu=0$ in Euclidean $\BR^4$, and define the state on surfaces
of constant $|x|$.  We then develop the canonical formalism with $|x|$
as time and quantize; the role of Hamiltonian is then played by
the dilatation operator $D$.
The conformal invariance prediction for the two-point
function
\eqn
\vev{\phi^i(x) \phi^j(0)} = {C_i \delta^{ij} \over |x|^{2\Delta_i}}
\enq
then (by spectral decomposition) implies that each primary field
$\phi^i(x)$ is associated with a distinct state $\ket{i}=\phi^i(0)\ket{0}$
of `energy' $D~\ket{i} = \Delta_i~\ket{i}$.

Superconformal invariance also leads to constraints on short
multiplets of supersymmetry.  The generic $\CN=4$, $d=4$ supersymmetry
multiplet with $256$ components must be
annihilated by all of the $S^{I\alpha}$.  This requires
the multiplet to belong to the appropriate representation of $SO(6)$ to
allow both sides of $\{Q,S\}=D+J$ (where $J$ are the $SO(6)$ charges)
to vanish.
Since $SO(6)$ charges are quantized, the dimensions and spectrum of short
multiplets are independent
of continuous parameters such as the coupling constant.
Thus ``chiral operators of $\CN=4$'' which create states in these multiplets
can be completely enumerated in perturbation theory.

One way to do this is to pick an $\CN=1$ subalgebra of $\CN=4$ as chiral
operators in $\CN=1$ are those which can be written as $\int d^2\theta$
in superspace; each $\CN=4$ multiplet will contain a unique sub-$\CN=1$
chiral multiplet of largest $U(1)_R$ charge.

The $\CN=4$ theory in $\CN=1$ superspace has three chiral adjoints $Z^i$
and a field strength $W^\alpha$, and superpotential $W=\Tr Z^1[Z^2,Z^3]$.
A representative set of chiral operators (not all) is
\eqn\label{operators}
O_{1n}^{i_1i_2\ldots i_n} = \Tr Z^{i_1} Z^{i_2} \ldots Z^{i_n}
\enq
Note that this set of operators is huge as we need to distinguish all
possible orderings of the indices $i_k$ in the large $N$ limit.
On the other hand we need to remove descendant operators such as those
predicted by the equations of motion,
\eqn
{\p W\over \p Z^i} = \epsilon_{ijk} [Z^j,Z^k] = D\bar D Z^i \sim 0
\enq
in terms of the chiral ring.  In other words, any operator in
(\ref{operators}) which includes a commutator is a descendant.

The result is that $\CN=4$ superconformal theory,
for any value of the Yang-Mills coupling constant, contains a sequence
of chiral operators $O_{1n}$ of dimension $\Delta = n$
which transform in the $n$-fold symmetric tensor of $SO(6)_R$.
The complete spectrum of chiral operators can be worked out \cite{ferrara}.

Note by contrast that we cannot make any statement about the spectrum
of non-chiral operators at strong coupling.  An operator such as
$\sum_i \Tr Y^i Y^i$ will have dimension $2$ in the free theory but
can gain an arbitrary anomalous dimension, presumably computable as
a power series in $g_{YM}^2$ and $N$.  When these corrections are large,
we have little direct control on these dimensions or indeed any
generic observable from the gauge theory point of view.

\section{Large $N$}

Another limit which is believed to simplify the gauge theory is that of
large $N$.  Although we can imagine different such limits, the best
studied (and probably best) is that of 't Hooft where we hold
the 't Hooft coupling $g_{YM}^2 N = \lambda$ fixed in the limit.
As is by now
classic (and reviewed in \cite{coleman}) the perturbative expansion in
this limit reduces to the sum of planar diagrams where a diagram with
$V$ vertices is weighed by the factor $(g_{YM}^2 N)^V$.
Furthermore,
corrections in $1/N$ are organized in a topological expansion,
with diagrams which can be drawn on a genus $g$ surface weighed
by $N^{2-2g}$.

This gives us a formal relation to string theory, in which
each operator written as a single trace of a product of adjoints corresponds
to an operator creating or destroying a single closed string, and
$1/N$ plays the
role of the closed string coupling constant $g_s$.
Specifically,
\eqn\vev{\Tr O_1} = N [O_1]_{disk}
+ N^{-1} [O_1]_{punctured\ torus} + \ldots
\enq
\eqa
&\vev{\Tr O_1 ~ \Tr O_2} =
 N^2 [O_1]_{disk} [O_2]_{disk} +
N^0 [O_1 O_2]_{annulus}  \cr
&\ + N^0 [O_1]_{disk} [O_2]_{punctured\ torus}
 + N^0 [O_2]_{disk} [O_1]_{punctured\ torus}
+ \ldots \label{factorization}
\ena
where each term $[O_1 \ldots]_{surface}$ corresponds to the contribution
of a string world-sheet with that topology and the specified operators
inserted at each boundary (or puncture).

This relation is very direct in
weak coupling perturbation theory, so when
this series converges, as was true for the ``old matrix models''
\cite{matrix}, we can confidently say that large $N$ field theory is
equivalent to a string.
We might hope to prove (\ref{factorization})
for general $g^2 N$ by analytic continuation.

However, we should recognize that the story
for most field theories is not so simple -- the weak coupling perturbation
theory is more typically asymptotic.  This leads to many potential
difficulties with the string interpretation.
The simplest appears in asymptotically free theories, where $g^2 N$ is
dimensionful, and the dynamically generated mass gap is known to
be non-analytic at $g^2 N=0$.  In the solvable example of $\CN=2$ SYM,
(\ref{factorization}) can be seen explicitly to fail \cite{ds}.

A different approach to string theory starts with a strong coupling
expansion around $g^2 N \sim \infty$.
(This is a subject with a long history; see \cite{dseries} and references
there.)
This can at present be made
precise only in two dimensions or on the lattice but in these cases
leads directly to a string with finite string tension, computable in
an expansion with finite radius of convergence.  However
these expansions generically predict large $N$ transitions and
a critical $g^2_c N$ below which the string expansion breaks down.

The situation may well be better in a superconformal theory and there
is no strong
argument against analyticity on the positive $g^2 N$ axis
in this case (but see \cite{li}).
One could then assume (\ref{factorization}) to obtain
a non-perturbative description of the theory,
as we explain in the last section.

\section{Near-horizon geometry and $\AdS_5$ supergravity}

The $p$-brane solutions were originally found as generalizations of
the extreme Reissner-Nordstrom solution of Einstein-Maxwell theory
and this leads us to ask whether the solutions have event horizons
and should be considered as black holes.  The story is different for
different solutions but what we need to do is consider the limit
$r=0$ and understand the behavior of the metric there.

This limit of the (single center) metric (\ref{threebrane}) is

\eqn\label{AdSmet}
ds^2 = \left(r^2\over \lambda^{1/2} l_s^2\right) dx^2
+ \left(\lambda^{1/2} l_s^2\over r^2\right)
	\left(dr^2 + r^2 d\Omega^2_5\right)
\enq
where we have written $y^i = r \hat y^i$; $\hat y^i$ is a unit vector
in $R^6$ parameterizing the sphere $S^5$, and $d\Omega^2_5$ is the
round metric on $S^5$.
This limit will be justified (we can drop the "$1$" term in
(\ref{threebrane}))
when
\eqn\label{curvature}
r << R \equiv \lambda^{1/4} l_s .
\enq

The first thing to notice is that the $r$ dependence cancels out in
the $S^5$ metric, leaving us with a solution $M \times S^5$.
Flux quantization for the five-form field strength tells us that
its restriction to $S^5$ must also be independent of $r$, while the
equation of motion tells us it will be a harmonic form on $S^5$.
The stress tensor for this special case is easily computed and
is invariant under $SO(4,1)\times SO(6)$, leading to a constant
curvature solution in both $S^5$ and in $M$.
The $SO(d-1,1)$ Lorentzian metric of constant negative curvature is
anti-de Sitter space $\AdS_d$ and in fact admits an action of $SO(d-1,2)$,
as we will see below.

Now that we have obtained $\AdS_5\times S^5$ as a near-horizon limit
we can also think in five-dimensional terms, by making the Kaluza-Klein
reduction of \IIb\ supergravity on $S^5$.
This was worked out in detail in \cite{vNw,gun1}; we summarize here.

According to the Kaluza-Klein program we must expand all the ten dimensional
fields in harmonics of the isometry group $SO(6)$ of $S^5$. In this way we
generate an infinite number of  $AdS_5$ fields
with spins ranging from $0$ to $2$. Each field has a definite $SO(6)$
content. One should then solve the linearized type \IIb\ supergravity
equations for these modes to determine the mass spectrum of the physical
states and their behaviour under the isometry group $SO(4,2)$ of $AdS_5$.
For the $AdS_5\times S^5$ solution the analysis has been caried
out in detail in  \cite{vNw}.

An alternative route to obtain information
about the spectrum is to use
the supersymmetry algebra $SU(2,2|4)$ of the $\AdS_5\times S^5$ background.
Let  $M_{AB}$, $A, B = 1,2,3,4$ or $0,-1$ denote the generators of its
$SO(4,2)$ bosonic subgroup of isometries of $\AdS_5$, and let
$B^M_N$ with $1\le M,N= 4$ denote generators of
$SU(4)\cong SO(6)$, the isometry group of $S^5$.
Supersymmetry generators correspond to solutions
of the Killing spinor equations (which set (\ref{susy}) to zero).
Half of these were discussed in section 2:
the D$3$-brane solution preserves an $SO(3,1)$
subgroup of $SO(4,2)$, generated by the subset $M_{ab}$, $0 \le a, b \le 3$,
and $16$ real supersymmetries, a $4$-plet of
Weyl spinors of $SO(3,1)$ which we denote by $Q^M$ with $M=1,..4$ a vector
of $SU(4)$.

The commutators of $Q^M$ with $K_a= M_{a 4}-M_{a,-1}$
now produce another $4$-plet of Weyl spinorial charges $S^M$, enlarging
the total number of unbroken supersymmetries to 16 complex.
The set $M_{AB}, B^M_N, Q^M, S^M $  generates   $SU(2,2|4)$.\footnote{
This group has an extra $U(1)$ factor which should be factored
out.\cite{gun2}}
As we commented in section 3 this symmetry group follows from combining
$\CN=4$ supersymmetry with the enlargement $SO(4,2)\supset SO(3,1)$ and
is also the superconformal symmetry of $\CN=4$ SYM.

Being the symmetry group of the background manifold,
the KK spectrum will be a discrete unitary (reducible)
representation of $SU(2,2|4)$.  We will discuss this representation theory
in some detail shortly.
The masses of $5$-dimensional modes will be given in terms of
the eigenvalues of
the generator $E=M_{0,-1}$. This operator generates translations
along a global timelike Killing vector field of $\AdS$ and therefore
is a useful choice for the $\AdS$ energy operator.

The simplest way \cite{gun1,gun2} to construct the lowest weight
unitary irreducible representation (UIR) of $SU(2,2|4)$ is to
introduce a set
of superoscillators $\xi^A=\{a^i, \alpha^\mu \}$ and
$\eta^M=\{b^r, \beta^x \}$, where
$i,  \mu, r$ and $x$ = 1,2. They satisfy the usual algebra of
the creation and annihilation
operators, viz,
$[ a^i, a_j]=  \delta^i_j$, $[ b^r, b_s]=  \delta^r_s$, $\{\alpha^\mu,
\alpha^\nu\}=  \delta^\mu_\nu$
and  $\{\beta^x, \beta^y \} = \delta^x_y $. All other commutators or
anticommutators are zero. Here
we have denoted the hermitian conjugate of $a^i$ by $a_i$, etc.
The $SU(2,2|4)$ generators are constructed as bilinears in these
oscillators.  For example the $16$ generators
$B^M_N$ of $U(4)$ are given by $\alpha^\mu \alpha_\nu, \beta^x \beta_y,
\alpha^\mu \beta^x$ and $\alpha_\mu \beta_y$.

The Fock space of the oscillators provides the vector space on which
one particular class of lowest weight UIR of $SU(2,2|4)$ are realized. These
representations are called the doubleton representions, because only one
pair of superoscillators
are used in their construction. Furthermore, although the doubleton
representations are not
part of the KK spectrum of the \IIb\ supergravity on the $\AdS_5\times S^5$
background, some of them can be
given an interpretation of massless  states in a conformal field thery in 4
dimensions. They
are massless because the entire set of doubleton representations
are in the kernel of the $4$-dimensional  mass operator
$m^2= P^a P_a$, \cite{gun2} where the  momentum
is defined by
$P_a= M_{a,0}+M_{a,-1},  a=0,1,2,3$.

The simplest doubleton representation is built using
the Fock vacuum  $|0>$ as the
lowest weight vector of $SU(2,2|4)$. It can be shown  that the multiplet
contains $2^4$ physical states:
$6$ real scalars, $4$ Weyl spinors and a vector potential, transforming
respectively in the vector, spinor and the singlet representations of $SO(6)$.
This is the $\CN=4$ super Yang Mills multiplet in $d=4$.

As another example we consider the direct sum of the two
doubleton representations built on the
lowest weight vectors $\xi\xi|0>$ and $\eta\eta|0>$. The resulting states
correspond to the
spectrum of the  $\CN=4$ superconformal gravity in $d=4$.
The full list of the doubleton representations have been given in
\cite{gun2}.

By taking tensor products of doubletons one builds other
repesntations of
$SU(2,2|4)$. An equivalent way to do this is to affix
an index $K=1, ...,p$ to the oscillators and write them as $a^i(K),
b^r(K), ...$.
 The commutators then become $[ a^i(K), a_j(K')]=  \delta^i_j
\delta_{KK'}$, etc.
The generators of $SU(2,2|4)$ are again bilinear invariants of the
$O(p)$ group acting on the index $K$. For example, the $U(4)$ generators are
 given by
$\alpha^\mu(K) \alpha_\nu(K), \beta^x(K) \beta_y(K),
\alpha^\mu(K) \beta^x(K)$ and $\alpha_\mu(K) \beta_y(K)$, where we sum over
the repeated $K$ label from $1$ to $p$.

By choosing different values of $p$ and different lowest weight spaces,
one can obtain many UIRs of $SU(2,2|4)$.
Each of these will contain
a finite number of UIRs of $SO(4,2)\times SO(6)$,
and each mass shell physical mode in $\AdS_5$
will come in such a UIR.
However, states in the KK spectrum
of  \IIb\ supergravity on  $\AdS_5\times S^5$
will correspond to using the Fock vacuum $|0>$ as the
lowest weight vector, and live in short multiplets of $SU(2,2|4)$.
This follows simply because KK reduction will only produce states
with spins not exceeding $2$.
Other choices of the lowest weight
vectors such as $\xi\xi...\xi|0>$ will produce a longer spin range and
thereby also longer multiplets.\footnote{For $p=2$ a complete classification
of these has been given in \cite{gun2}.}

Restricting attention to this choice,
the multiplets are conveniently characterized by their lowest weights
$(J_L,J_R, E)$ under the subgroup $SO(4)\times SO(2)\subset SO(4,2)$.
For example, for each $p>1$, the $SO(4,2)$ representations defined
by the lowest weight vector
$(1,1,2p+4)$ contain all spin $2$ modes in
$AdS_5$ transforming in the representations  characterized by the Dynkin
label $(0,p-2,0)$ of $SO(6)$.
The complete multiplet contains $128 {{p^2(p^2-1)}\over 12}$
fermionic and the same number of bosonic states
(see Table I of \cite{gun1}).

For $p=2$ the spin $2$ mode is the
$5$-dimensional graviton which is a singlet
of $SO(6)$.  In addition this multiplet (the massless supergraviton
multiplet)
contains four complex gravitini, $15$ vector fields in the adjoint
of $SO(6)$, plus spin zero and spin $1/2$ objects
for a total of $128+128$ physical states.

The $p>2$ representations will correspond to the massive KK towers.
In fact, the complete KK spectrum on $\AdS_5\times S^5$ is obtained
by taking a single copy of each $p\ge 2$ representation.
Again with each $SO(4,2)$
lowest weight we can associate a particular mass shell mode in $\AdS_5$
in a definite $SO(6)$ representation.

To close this section we remark that
since the supersymmetries transform under $SO(6)$, so will the gravitinos.
The presence of vector fields in the adjoint of $SO(6)$ thus means that
the $5$-dimensional theory is one of the known gauged supergravities
\cite{diverse}.

\section{More on AdS}

The $SO(d-1,2)$ symmetry of $\AdS_d$ is
directly analogous to the $SO(d+1)$ action
on $S^d$ and can be realized linearly in the same way, by embedding
as a surface in one higher dimension.  Thus we introduce
a $d+1$ dimensional flat space with a
Lorenzian metric of signature $(2,d-1)$ and coordinates $X^A$,
and represent $\AdS_d$ as the surface satisfying
\eqn\label{constraint}
X^2_{-1} + X^2_{0} - \sum_{i=1}^{d-1} X^2_{i} = R^2.
\enq
with the curvature radius $R$ given in (\ref{curvature}).
This construction also provides global coordinates on $\AdS_d$:
the relation between the D-brane coordinates $(x^\mu,r)$ and the $X^A$ is
\eqa
r &= X_{-1} + X_4 \cr
x^\mu &= {X^\mu R \over r} .
\ena
We see that the full metric is non-singular for $r\le 0$.
This region is behind an event horizon from the point of view of
an asymptotic observer in the original D-brane metric (\ref{threebrane}).

The global structure of the AdS metric is better visualized by solving the
constraint (\ref{constraint}) in a different way (take $R=1$):
\eqa
X_{-1} &= {\cos t \over {\cos \rho}} \cr
X_0 &= {\sin t \over{\cos \rho}} \cr
X_i &= z_i tan \rho \cr
& \sum_i z_i^2 = 1
\ena
which leads to the metric in the form
\eqn
ds^2 = {1\over\cos^2 \rho} \left(-dt^2 + d\rho^2\right)
+ \tan^2\rho\ d\Omega^2_{d-2} .
\enq

\begin{figure}
\epsfxsize2.5in\epsfbox{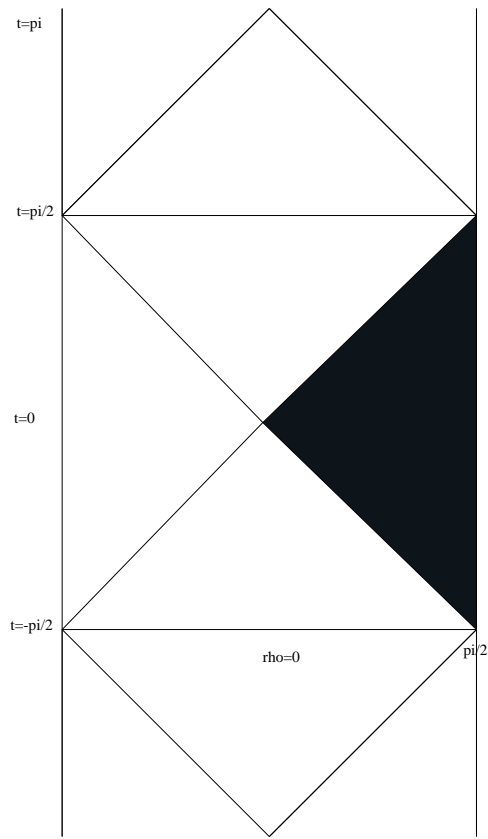}
\caption{Penrose diagram for AdS space-time}
\end{figure}
These are usually called `global' coordinates as they cover all of AdS
as a single copy of the region $0 \le t \le 2\pi$, $0 \le \rho \le \pi/2$
and $z_i \in S^{d-2}$.  (For $d=2$ we take $S^0$ as the two points
$z^1=\pm 1$).  Furthermore, the metric is conformally flat in the $\rho t$
plane, so the causal structure is easy to understand, and is depicted
in Figure 1.  The shaded region is the region described in the original
(`Minkowski') coordinates.

Note that the coordinate $t$ is periodic and thus AdS space
contains closed time-like curves.  Often these are problematic and
to avoid this we can work instead on the universal cover by simply
ignoring the periodicity of $t$.  If we do so, the future boundary of the
shaded region is a horizon, but only because the infalling observer
is forced to leave through a copy
of the original asymptotic region.  The physics is quite different from that
of a black hole horizon.

Rather, the striking feature of AdS geometry is that
the limit $r\rightarrow \infty$
(equivalently $\rho\rightarrow \pi/2$) takes us to a {\it time-like}
boundary, with geometry (on the cover) $S^{d-2} \times \BR$.

The presence of the timelike boundary leads to major differences with
physics in Minkowski space.  For one thing, we do not have a well-posed
initial value problem unless we put boundary conditions there.
Of course we always need boundary conditions at infinity,
but for Minkowski space the boundary conditions can be decomposed into
incoming and outgoing waves, leading to the usual idea of particle
and S-matrix.

By contrast in AdS a small fluctuation can typically be
decomposed into normalizable and nonnormalizable components;
the correct way to proceed in such situations \cite{SeiShen,Bala}
is to quantize the normalizable fluctuations
(each such mode is associated with a creation and annihilation operator)
while the nonnormalizable modes are instead considered to be a static
background.  The two types of modes are distinguished by their asymptotics
as $\rho\rightarrow \pi/2$ and so the background can
be determined from boundary conditions at $\rho = \pi/2$ uniquely
up to a choice of normalizable component.

Thus we can regard
all observables in $\AdS_d$ quantum gravity as functions of a set of $d-1$
dimensional fields, in one-to-one correspondence with non-normalizable
modes and thus with fields on $\AdS_d$.

\section{The AdS/CFT correspondence}

As we saw in the previous sections,
superstring theory admits a class of BPS solitons --
N parallel coincident D$3$-branes --
with two very different descriptions in the near-horizon limit $r << R$.

On the one hand, we have the dynamics of small fluctuations of \IIb\ %
supergravity around the $\AdS_5 \times S^5$ background.  In general
this is only an approximation to \IIb\ string theory.
First of all, we need $g_s << 1$ to justify ignoring quantum
corrections\footnote{
We can treat the S-dual limits $g_s \rightarrow \infty$ and
so on by first applying \IIb\ S-duality before making our analysis.
We would reach the same conclusions in terms of the S-dual coupling
which is going to zero.}
(which are ill-defined in pure supergravity).  Furthermore we have
stringy corrections -- in an effective
Lagrangian language, we know that higher order corrections exist at
leading order $l_s^6 R^4$ in the pure gravity sector, and presumably
at all higher orders in $l_s$.  However when the background curvature
radius $R >> l_s$ and all energy scales of interest satisfy $E l_s << 1$,
\IIb\ supergravity is a good approximation.  Using (\ref{curvature})
this corresponds to
\eqn
g_s N >> 1 \qquad {\rm and} \qquad g_s << 1
\enq
so evidently we require $N >> 1$.

On the other hand, for $g_s N << 1$ we are more tempted to regard the
D$3$-branes as a small perturbation of flat space.  From this point of
view the full dynamics is the open strings attached to the D$3$-branes,
coupled to the closed strings in the bulk.  If we then restrict attention
to the near-horizon limit $r << R$, since we have $R << l_s$ in this
regime we are in the substringy regime $r << l_s$ and can restrict
attention to the lightest open strings, in other words $N=4$ SYM coupled
to supergravity.

Actually, if $g_s N << 1$, this is a redundant description.  As was argued
in \cite{dkps}, by considering
world-sheet duality as it works in the perturbative string expansion,
all closed string effects, including supergravity,
in this limit are actually contained
in the dynamics of the open string theory.  On the other hand, supergravity
predictions will obtain large corrections due to the massive closed string
modes.

The upshot of all this is that we have two descriptions of the $3$-brane
which appear to have non-overlapping regimes
of validity.

We can now consider the question:
What happens when we extrapolate this second, gauge theory description
to the regime $g_s N >> 1$ ?  As long as we consider low energies and
$r << l_s$, all of the conditions for the gauge theory description
appear to be met.  But we can also fulfill the condition for the
applicability of supergravity, $R >> l_s$.  Could it be that both
descriptions are simultaneously valid ?  If so, could it be that
we can interpret the gauge theory at finite $g_s N$ as \IIb\ string
theory on AdS with the $l_s$ corrections included ?

As we saw in detail, the two descriptions have a lot in common --
for example, the symmetry groups are the same.
However the crucial remaining ingredient in making such a claim is that
we need to argue that the description involving gauge theory coupled
to supergravity is also redundant for $g_s N >> 1$ -- that the
gauge theory includes all of the states of supergravity in this regime as
well.

Using the operator-state correspondence of the gauge theory,
we can rephrase this
as follows: if there is a one-to-one correspondence between
the operators of $d=4$, $\CN=4$ SYM and the states of the
$AdS_5\times S^5$ compactification of type \IIb\ theory
(including their transformation properties under $SU(2,2|4)$),
then gauge theory must contain the states of supergravity already --
we do not need to add them as separate degrees of freedom.
Since the large $N$ limit at fixed $\lambda$ corresponds to weak string
coupling, we expect this to work for single string states: that these
are in correspondence with single trace operators in the gauge theory.

Without results for the gauge theory at large $\lambda$, we can only check
this for the short multiplets, whose dimensions are unrenormalized.
Happily all the states of this KK supergravity lie in short multiplets
and the check for these states works \cite{witten,ferrara}.
{}From the arguments above we might regard this as proof; a skeptic might
reply that the one-to-one correspondence for these operators was also
required by the gauge theory coupled to supergravity picture,
which already required
a gauge theory current to couple to each supergravity field.

We can extend the conjecture to any $N$ and $\lambda$ and in the
large $N$ limit we should make contact with the classical (genus zero)
limit of the \IIb\ theory.  The (Polyakov) bosonic
string action in the metric (\ref{AdSmet}) is
\eqa
S &= {1\over 2\pi l_s^2} \int d^2\sigma\ g_{\mu\nu}(X) \p X^\mu \p X^\nu \cr
&= {1\over 2\pi}
\int d^2\sigma\ \lambda^{1/2}{(\p u)^2\over u^2}
+ \lambda^{-1/2} u^2 (\p x)^2 \cr
&\qquad  + \lambda^{1/2} g_{S^5}(\hat y) (\p \hat y)^2 .
\ena
After substituting $u=r/l_s^2$, we can take the limit $l_s\rightarrow 0$,
and still get a non-trivial closed string theory, whose action depends on
$\lambda^{1/2}$, the curvature radius squared (of both AdS and $S^5$)
in string units.
Stringy corrections to supergravity results
will be naturally computed as a series in
$1/\lambda^{1/2}$.  The superstring action also includes the fermions
and their coupling to the background RR field.  Various forms of this
action have been proposed \cite{tseytlin} though
as yet none have been quantized.

In a way the case for this stronger conjecture is better -- we know there
is more than one candidate classical
theory with $SU(2,2|4)$ symmetry and this spectrum
of short multiplets (indeed taking different $\lambda$ produces different
theories) but if we could assert that {\it all} such theories fall into
this one parameter family the equivalence would be proven.
This type of assumption underlies all duality arguments but we should
recognize that except in the most intensively studied cases (e.g.
supersymmetric gauge theory) it is at some level an argument from ignorance,
especially because we cannot assume the theory is local in
ten-dimensional space-time.
Thus it is worthwhile to look for further, concrete tests of the conjecture.

For finite $\lambda$, we now have massive string states as well
with masses determined by the string tension and these would correspond
to gauge theory operators with dimension $\lambda^{1/2}$ \cite{gkp}.
The simplest
picture is thus that all nonchiral gauge theory operators acquire such
an anomalous dimension at large $\lambda$, a prediction which remains to
be tested.

A more general statement of the conjecture can be made by considering the
possibility of non-trivial boundary values of the fields in AdS, as
we explained in section 6.  It is quite natural to associate these with
non-trivial {\it couplings} to the corresponding operators in the gauge
theory.  We then can conjecture the equivalence of the complete generating
functionals
\eqa
\int [DA,Y,\chi]\ e^{-{N\over\lambda}S_{N=4} + \sum_i \int d^4x~ \hat\phi_i(x)
O_i(x)}
\cr
= \int_{\lim_{r\rightarrow\infty} \phi_i(x,r) \sim \hat\phi_i(x)}
[Dg,\phi,\ldots]\ e^{-S_{\IIb}}
\ena
in the two quantum theories.
In the large $N$ limit, the \IIb\ side reduces
to genus zero string theory which is dominated by a single ``master field;''
in the large $\lambda$ limit this reduces to \IIb\ supergravity.
Thus any gauge theory correlation function can be computed in this limit
by a corresponding computation in classical field theory.  For more details
on this we refer to \cite{witten}.

Perhaps the most striking result which has been obtained in this direction is
\cite{minwalla,freedman} in which three-point functions of chiral operators
are compared and found to agree at large $N$ in the two limits of large
and small $\lambda$.

\section{Wilson loops}

Quite a lot of work has been done on the correspondence following
\cite{maldacena}\ and we will not even try to summarize it all here.
However as of this writing it seems fair to say that while
some striking
agreements and no definite contradictions have been found,
no convincing microscopic argument
has been made for why it should work.

By a microscopic argument we mean some identification in one limit
(say, perturbative gauge theory) of the degrees of freedom which
control the other limit, as we have in the earlier examples of superstring
duality.  For example, the D$0$-branes of \IIa\ superstring theory
clearly become the lightest BPS states in the strong coupling limit,
leading to the conjecture that this is eleven dimensional
supergravity;
furthermore we have a very explicit definition of the D$0$-branes in
the weak coupling limit, which we could in principle extrapolate to
define the supergravity, as in \cite{bfss}.
Making this extrapolation in practice is
difficult as it seems to require understanding bound states of large
numbers of D$0$-branes, but at least we know the bound states exist.

In the present case, the analogous object would seem
to be the \IIb\ string in the $N=4$ gauge theory, and a convincing argument
that the string is present in the gauge theory would remove any doubt.
Although it has been repeatedly conjectured over
the years that gauge theories contain strings in the large $N$ limit,
based on the formal correspondence we described in section 4,
all attempts to make this concrete so far have run into problems.

All attempts we know about have identified the Wilson loop operators in
the gauge theory as string creation operators and so we should start
by making sense of this identification for $\AdS_5 \times S^5$.
Now a Wilson loop in gauge theory is a four-dimensional object and
to formulate an operator which creates a superstring in the nine-dimensional
boundary of $\AdS_5 \times S^5$ we clearly need to include the operators
$Y^i$ as well.  In general we also need to specify the spin state
of the string and thus insert operators $\chi$;
however this should not affect the leading results in an expansion in the
inverse string tension $1/\lambda^{1/2}$, which are plausibly given just
by minimizing the Nambu-Goto action in $\AdS_5 \times S^5$ for a world-sheet
with the specified boundary.  We thus arrive (for large $\lambda$)
at the conjecture
\cite{maldawilson}
\eqn\label{mloop}
{1\over N}\vev{\Tr P e^{\int_L iA + \theta\cdot Y}}_{g.t.}
= e^{-S_{N.G.}|_{\partial \Sigma = L}} .
\enq
Let us consider a boundary at a single point in $S^5$ and a rectangular
Wilson loop $L$ in $\BR^4$ with two sides at distance $\Delta X^1 =L$
extending for a time $\Delta X^0 = T$.  This loop will be spanned
by a world-sheet which in the large $T$ limit will be independent
of $X^0$, so the amplitude will be determined by a static potential,
$\exp -T V(L,\lambda)$.
The string can then be described
by world-sheet coordinates $\tau=X^0$ and $\sigma=X^1$; its
embedding in AdS is determined by $u(\sigma)$.
The Nambu-Goto action then reduces to
\eqn
S_{N.G.} = \lambda^{1/2} T \int d\sigma \sqrt{ u^2 +
{1\over u^2}\left({\p u\over \p\sigma}\right)^2 } .
\enq
Minimizing this action is a solvable problem \cite{maldawilson}
and the result is
\eqn
V \equiv {\kappa(\lambda)\over L}
= {4\pi^2 \lambda^{1/2}\over \Gamma({1\over 4})^4 L}
\enq
Conformal invariance guaranteed that the
amplitude would be a function of $T/L$ and thus the non-trivial
information is in the effective coupling $\kappa(\lambda)$.

We can compare this to the gauge theory at weak coupling,
at least on a qualitative level.
There the leading contribution comes from exchange
of a single gluon or scalar, giving
\eqn
\kappa = {\lambda\over 2\pi} + \ldots
\enq
We can easily imagine a continuous, monotonic function $\kappa(\lambda)$
with these two asymptotic behaviors, giving us a consistent picture.

{}From the \IIb\ string point of view,
the dependence $\kappa \sim \lambda^{1/2}$ at large $\lambda$
comes just because the string tension $T_s \sim 1/l_s^2$.
On the other hand it
is quite surprising from the gauge theory point of view, so confirming
this would provide strong new evidence for the conjectured string.
However, no effective techniques for this gauge theory computation
exist at present.

A natural starting point which takes the large $N$ limit but keeps
$\lambda$ finite
is the ``loop equation'' approach, first formulated for pure YM
theory by Migdal and Makeenko \cite{migdal}.
Let
\eqn
W(L) \equiv P e^{i\int_L A}
\enq
so that the complete set of Wilson loop operators is $\Tr W(L)$ for
all loops $L$ starting and ending at a point $x$.
(This is in pure gauge theory;
adding the additional fields of SYM and including loops such as (\ref{mloop})
is easy at this formal level.)

The starting point is the Schwinger-Dyson equations,
\eqa
0 &= \int [DA]\ \Tr~\left({\delta\over\delta A_\mu(x)} W(L)\right)
e^{-{N\over\lambda}\Tr F^2} \cr
&=  \vev{ \sum_{L_1L_2 = L} \Tr W(L_1) \Tr W(L_2) }\cr
&~~~ - {N\over\lambda} \vev{\Tr W(\nabla^\nu F_{\mu\nu} W(L) } .
\ena
Using techniques described in \cite{migdal}, the second term can be
rewritten as a ``differential operator on loop space'', a difference
of $W(L')$ and $W(L)$ where $L'$ is obtained from $L$ by adding infinitesimal
pieces of loop at $x$.
The large $N$ limit is then taken by using (\ref{factorization}) on the first
term and keeping only the $O(N^2)$ part.  The final result is a closed
equation for the $O(N)$ expectation values of all Wilson loops (or
``master field''),
\eqa
\vev{\Tr W(L')} - \vev{\Tr W(L)} = \cr
{}~~~ \lambda \sum \vev{\Tr W(L_1)} \vev{\Tr W(L_2)} .
\ena
Setting the left hand side to zero would just be the classical equation of
motion; if $L$ intersects itself at the point $x$
we also obtain a quantum source term which is a sum over all ways of
intercommuting the loop $L$ at $x$.

All this is precise for a regulated gauge theory and one can argue that
area law is a natural ansatz which solves it (this sounds good for pure YM);
furthermore one can reproduce conventional weak coupling perturbation
theory by expanding the $W(L)$'s in powers of the fields.
However its non-linearity makes direct contact with
string theory -- e.g. checking whether a concrete first quantized
string wave functional $\psi[X(\sigma)]$ is a solution --
quite difficult.

Furthermore we know no regulated form of pure $\CN=4$ SYM which preserves
the supersymmetry, so using this probably requires postulating a
renormalized form of the equation.  In addition we run into all of the
old problems of QCD string  \cite{qcdstring}
such as the great difference between a loop
functional $W[L]$ of continuous loops $L$, and all known superstring
loop functionals $\psi[X(\sigma)]$ which have support on {\it discontinuous}
loops.

Nevertheless, now that we have a precise and
better motivated conjecture for the appropriate string in this case,
we can hope that progress along these lines will be made in the near future.

\end{document}